\documentclass[reqno]{amsart}
\usepackage{oldlfont}
\usepackage{amssymb}
\usepackage{graphicx}

\voffset = - 1 cm
\hoffset = - 1.5 cm
\newcommand{\bra}[1]{\langle #1 \! \mid}
\newcommand{\ket}[1]{\mid \! #1 \rangle}
\newcommand{\braket}[2]{\langle #1 \! \mid #2 \! \rangle}

\begin{document}
\title[The Shape of Spatial  Atoms]{Quantum Geometry Phenomenology: Angle and Semiclassical States}
\author{Seth A. Major}
\date{December 2011}
\address{Department of Physics, Hamilton College,
Clinton NY 13323 USA}

\begin{abstract}
The phenomenology for the deep spatial geometry of loop quantum gravity is discussed.  In the context of a simple model of an atom of space, it is shown how purely combinatorial structures can affect observations.  The angle operator is used to develop a model of angular corrections to local, continuum flat-space 3-geometries.     
The physical effects involve neither breaking of local Lorentz invariance nor Planck scale suppression, but rather reply on only the combinatorics of $SU(2)$ recoupling.  Bhabha scattering is discussed as an example of how the effects might be observationally accessible.
\end{abstract}

\maketitle

\section{Introduction}

Due to the extreme physics defined by the Planck scale, quantum gravity phenomenology is only possible with ``lever arms". For instance, when local Lorentz symmetry is violated the lever arm of bringing the Planck scale within reach of observation is the magnificent sensitivity of particle physics in the effective field theory framework to the breaking space-time symmetries \cite{jlm,limits,jlm_rev,lm_rev}. This paper reviews the lever arm suggested in \cite{obs_angle} and contrasts this model with semi-coherent states peaked on discrete classical geometries in flat space. The model's lever arm is intrinsic to the discrete geometry of an atom of spatial geometry. As such it is an example of a phenomenology arising not from breaking local Lorentz invariance but rather from the fundamentally combinatoric description of spatial geometry. The combinatorics of an atom of geometry, the spin network node, suggests a dimensionless ``shape" parameter that serves as an expansion parameter for the corrections to classical, flat-space continuum angular geometry. An operational view is taken in which (angle) measurements determine the local deep geometry and which contrasts with results from semi-coherent states that are peaked on (oriented) classical polyhedra. The extent to which local geometry can be re-constructed from observational data is discussed.

\section{Angle Operator}
\label{intro_angle}

The angle operator, originally defined in \cite{angle}, may be conveniently cast into the combinatorial framework of loop quantum gravity. The angle operator is defined at a node.  Links incident to $n$  are partitioned into three sets $C_{1}$, $C_{2}$, and $C_3$.  One may visualize the partitioning as arising from three regions in the surface dual to the node.  However, the combinatorics only requires a tripartite partition. Three gravitational field operators $L^i_{1}$, $L^i_{2}$, and $L^i_{3}$ are defined by these partitions.  In terms of these field operators the quantum angle operator between dual surfaces corresponding to partitions $C_1$ and $C_2$ is
\begin{equation}
	\label{Aop}
\hat{\theta}_{(12)} := \arccos \frac{L^i_{1} L^i_{2}}
{|L_{1}| \, | L_{2} |},
\end{equation}
in which $|L| = \sqrt{L^{2}}$.  The sum $\sum_{k=1}^3 L^i_k$ vanishes due to gauge invariance and exhaustive partitions. This partitioning of links incident to $n$ gives a preferred intertwiner basis.  These are given by trivalent tree graphs that connect in an ``intertwiner core".  Deriving the spectrum of the angle operator of equation (\ref{Aop}) is a simple exercise in angular momentum algebra  \cite{angle}.
\begin{equation}
\label{angle_spectrum}
\begin{split}
\hat{\theta}_{(12)} \ket{j_1 \, j_2 \, j_3} &= \theta_{(12)} \ket{j_1 \, j_2 \, j_3}  \text{ with } \\
\theta_{(12)} &= {\rm arccos} \left( \frac{j_{3}(j_{3}+1) -
j_{1}(j_{1} +1) - j_{2}(j_{2}+1)} {2 \left[ j_{1}(j_{1} +1 ) \, 
j_{2}(j_{2} +1 ) \right]^{1/2}} \right).
\end{split}
\end{equation}
There are two aspects of the continuum angular distribution that are hard to model, small angles are sparse and the distribution of values is weighted toward large angles. The asymmetry persists even when the spins are very large \cite{mikes,major_seifert}. 

It is convenient to visualize the action of the angle operator on classical polyhedra, with faces dual to the incident links. The closed dual surface of the node, topologically $S^{2}$, is partitioned into three regions, $S_{1}$, $S_{2}$, and $S_{3}$, such that all the regions $S_{k}$ are simply connected. The intertwiner core then represents a decomposition of the polyhedron with ``internal areas" determined by $\vec{n}$. Although this picture is convenient, states may be deeply quantum and so our intuitive classical notions of face adjacentcy and dihedral angles may not apply.  In the Discussion I ask how much of this classical information semiclassical states can determine.

The notation is as follows. Twice the sum of the representations on the links incident to the node in partition $C_k$ is denoted by the ``flux"  $s_k$ also denoted $\vec{s}$.  In the dual surface picture this is the flux of spin through the respective surfaces, roughly the face areas.   The quantities $n_k= 2 j_k$, the internal areas, uniquely specify the intertwiner core.  The core states are $\ket{\vec{n}}$. The fluxes $s_k$ and core labels $n_k$ are distinct and satisfy $n_k \leq s_k$.

\section{Combinatorial phenomenology}
\label{phenom}

It was apparent in the numerical studies of \cite{mikes} that the asymmetry in the angular spectrum shifted the distribution for $\vec{n}$ at fixed $\vec{s}$, $p_{\vec{s}}(\vec{n})$, away from the usual $\sin \theta$ distribution of angles in three dimensional flat space. To recover this it was necessary in these studies to take large fluxes, and, in particular $1 \ll s_j \ll s_3$, $j=1,2$. Fluxes $\vec{s}$ that satisfy these relations are called ``semi-classical fluxes".

There is another reason why the we might wish to consider nodes with large spin.  Of course most physical processes we currently consider, such as scattering events, are ``local" on the scale of the theory being tested. But in terms of the quantum geometry the scales are very large, typically many orders of magnitude above the Planck scale.  In the volume operator likely to be relevant for the combinatorial framework, the volume scales as the $(\text{total flux})^{3/2}$. The scaling with volume can be used to define an effective length $\ell_s$ and an effective energy $M_s = M_{Pl}/\sqrt{s}$. A surprising result in \cite{mikes,major_seifert} suggests -- but does not predict -- that to model the correct distribution of angles in 3-space, the total fluxes are $10^{32}$ giving an effective length scale of  about $10^{-19}$ m, a perhaps not altogether hopeless scale.

To model an atom of 3-geometry suppose that the probability measure on the space of intertwiners is uniform. Further assume that all incident links to the node are monochromatic and ``simple", are spin-$\tfrac{1}{2}$. 
The combinatorics of the model can be solved analytically for semi-classical fluxes. For large flux $\vec{s}$ the normalized probability distribution is given by 
\begin{equation}
\label{ndist}
p_{\vec{s}}(\vec{n}) \simeq \prod_{i=1}^3 \frac{n_i}{s_i}  \exp \left( -\frac{n_i^2}{2s_i} \right).
\end{equation}
Accessible measurements of the atom include 3-volume, (roughly) determined by the total flux, and angle, determined by the states $\ket{\vec{n}}$ of the intertwiner core.  The fluxes $\vec{s}$ determine a mixed state,
\begin{equation}
\label{mixedstate}
\rho_{\vec{s}} = \sum_{\vec{n}} p_{\vec{s}}(\vec{n}) P_{\ket{\vec{n}}}
\end{equation}
where $P_{\ket{\vec{n}}}$ is the projector on the orthonormal basis of the intertwiner core. The sum is over the admissible 3-tuple of integers $\vec{n}$ such that $n_i \leq s_i$. In the discrete case the projector is $P_{\ket{\vec{n}}} = \ket{\theta_I} \bra{\theta_I}$ where $\ket{\theta_I} = \sum_{\vec{n}} c_{\theta_I} (\vec{n}) \ket{\vec{n}}$. 

The probability of finding the angle eigenvalue $\theta_I$ in the mixed state $\rho_{\vec{s}}$ is 
\begin{equation}
\label{discrete_dist}
\text{Prob}(\theta = \theta_I ; \rho_{\vec{s}}) = \text{tr}\left( \rho_{\vec{s}} P_{\theta_I} \right)
= \sum_{\vec{n}} p_{\vec{s}}(\vec{n}) |\braket{n}{\theta_I} |^2 \equiv p_{\vec{s}}(\theta).
\end{equation}
This procedure can be used to calculate $p_{\vec{s}} (\theta)$ in the continuum approximation. For semi-classical fluxes and a value of the measured angle $\theta$, now taking continuous values, within an interval $\Delta \theta = (\theta-\delta \theta,\theta +\delta \theta)$ the geometric probability distribution is
\begin{equation}
\text{Prob}(\theta \in \Delta \theta; \hat{\rho}_s) = \text{tr} \left( \hat{\rho}_s \hat{E}_{\Delta \theta} \right) = \int d^3n P_{\vec{s}}(\vec{n}) \bra{\vec{n}} \hat{E}_{\Delta \theta} \ket{\vec{n}}
\end{equation} 
where $\hat{E}_{\Delta \theta}$ is the projector onto the interval $\Delta \theta$. Geometrically it projects the state onto a (thickened) surface in $\vec{n}$-space given by $\theta(n) \in \Delta \theta$.  Taking the limit $\delta \theta \rightarrow 0$ gives continuum approximation to equation (\ref{discrete_dist})
\begin{equation}
\label{rawdisttheta}
P_{\vec{s}}(\theta)  := \int d^3n  \, p_{\vec{s}}(\vec{n}) |c_\theta(\vec{n})|^2 \delta \left( \theta - \theta(n) \right). 
\end{equation}
The integration of equation (\ref{rawdisttheta}) is straightforward \cite{obs_angle}.  The key step in the calculation is the identification of the ``shape parameter" $\epsilon := \sqrt{s_1 s_2}/{s_3}$ that measures the asymmetry in the distribution of angles. 

The resulting measure, when expressed in terms of Legendre polynomials and to $O(\epsilon^3)$, is \cite{obs_angle}
\begin{equation}
\label{moddist}
\rho_\epsilon (\theta)  \simeq \sin \theta \left( 1 - \frac{8}{\pi} P_1(\cos \theta) \epsilon  + \frac{3}{2} P_2(\cos \theta) \epsilon^2 \right).
\end{equation} 
The affect of the modified distribution of polar angles is that the `shape' of space is altered by the combinatorics of the vertex; the local angular geometry differs from flat 3-space. While these effects would be in principle observable at any flux, the results here are valid for semi-classical flux,  $1 \ll s_j \ll s_3$ for $j=1,2$.  In this model the total flux $s = \sum_i s_i$ determines the 3-volume of the spatial atom and thus an effective mesoscopic length scale, $\ell_s = \sqrt{s} \ell_P$, greater than the fundamental discreteness scale of $\ell_P$. The combinatorics of the intertwiner provides a lever arm to lift the fundamental scale of the quantum geometry up to this larger mesoscopic scale. So while the the shape parameter $\epsilon$ is free of the Planck scale, the effective length scale, determined by total flux $s$, is tied to the discreteness scale of the theory.

\section{Example: Scattering}
\label{scattering}

If the scale $\ell_s$ of the spatial atom is large enough then the underlying geometry would be accessible to observations of particle scattering. To see how the combinatorial effects might be manifest I'll briefly discuss  combinatoric corrections to Bhabha scattering. This process is convenient because the $e^+e^-$ scattering process involves ``point-like" fundamental particles and for the practical reason that data is readily available. We are used to using scattering data to learn about the effective field theory at a certain energy scale.  But the angular data is also a measurement of (much more) local geometry on some effective scale. The modification of the local angular geometry of space - the shape - leads to a modified solid angle. This is the dominant effect.

Short distance modifications to QED may be expressed in the Drell parameterization \cite{drell}
corresponding to a short range potential added to the Coulomb potential.  The local, discrete geometry manifests itself through the combinatorial corrections to the local geometry.  For instance the Bhabha scattering cross section is 
\begin{equation}
\left( \frac{d \sigma}{d \Omega} \right) / \left( \frac{d \sigma}{d \Omega} \right)_{QED} = 1 \mp \left(\frac{3 s}{\Lambda^2_\pm} \right) \left( \frac{ \sin^2 \theta }{3 + \cos^2 \theta} \right) \left( 1 + \frac{8}{\pi} \cos (\theta) \epsilon + \dots \right).
\end{equation}
It remains to be seen whether a derivation along the lines of \cite{drell} that incorporates shape corrections would yield this form of the correction.  A comparison between the model and the data (see \cite{obs_angle}) shows that the shape correction reduces power at small angles and increases power at large angles.

\section{Discussion}
\label{concl}

This paper reviews a new quantum gravity phenomenology, one that explores effects arising from combinatorial structures in the deep spatial quantum geometry of LQG \cite{obs_angle}.  The example model relies on the combinatorics of a specific discrete model of spatial geometry, that of a single atom of spatial geometry, the spin network node. This model shows that potentially observable effects of quantum geometry need not be tied to (obvious) violations of local Lorentz symmetry and that a scale above the fundamental scale of the theory can arise out of combinatorial effects. In the context of an atom of space, the underlying combinatorics may be enough to determine corrections to flat, continuum 3-geometry.

We can compare this model to currently-employed coherent states, such as the semi-coherent states of Levine and Speciale \cite{LS}, peaked on geometry and spread on curvature.  These group averaged states $\ket{\underline{j} \underline{\hat{n}}}$ are built from $SU(2)$ coherent states and are peaked on states satisfying closure at large spin.  They are thus peaked on classical (oriented) geometries at large spin.  Relatively moderate ``large" spins are required, e.g. $j \sim 100$.  The basic coherent states are constructed from $N$ spins (or face areas) and $N$ directions. These directions are preserved in the averaging
\begin{equation}
\frac{\bra{ \underline{j} \underline{\hat{n}} }  \cos \theta_{ij}  \ket{ \underline{j} \underline{\hat{n}}}}
{\braket{\underline{j} \underline{\hat{n}}}{\underline{j} \underline{\hat{n}}} }
\simeq \hat{ n}_i \cdot \hat{n}_j
\end{equation}
at (moderately) large spin.  Thus the LS states, constructed from $3N-3$  degrees of freedom, reproduce classical angles in 3-space at moderate flux, as long as $N$ is large and the classical directions $\underline{\hat{n}}$ are chosen uniformly.  In this case the LS states exhibit no lever arm.  Bianchi {\em et. al} show in \cite{poly} that the geometry of a classical polyhedron with $N$ faces is determined by $3N-6$ data.  The LS states are specified with $3N-3$ degrees of freedom.  Due to the peakedness of the states. It is clear that the distribution following from the LS states is classical in the large spin limit, if the $\hat{n}$ directions are chosen uniformly.  Further, the relevant ``large" spins are many orders of magnitude less than those in that suggested a possible combinatoric lever arm.  The classical geometry of an (oriented) polyhedron is approximated well for relatively modest spins and volume.
   
Taking an operational point of view the accessible measurements from scattering are an effective volume and a set of angles.  From the spin geometry theorem \cite{penrose,moussouris} we know that classical angles in Euclidean 3-space exist as long as the relative uncertainties in angle are small (the state is ``$\delta$-classical"). Suppose for the moment that the mesoscopic quantum geometry is homogeneous.  In the course of a scattering experiment angles and uncertainties are measured.  Two questions arise: Can the resulting angular distribution data be described as a set of angles of a classical $N$-sided polyhedron?  What sort of corrections might we expect for an angular semi-classical state? From the work of \cite{poly} we know that the answer to the first question is negative: Lacking the face areas, angular data are too few to uniquely specify a classical polyhedron geometry.  To answer the second question, we need to specify a (class of) closed coherent states peaked on angles, not on classical polyhedra with a fixed number of faces.  From the spin geometry theorem we know that angles in Euclidean 3-dimensional space may be approximated to arbitrary accuracy for $\delta$-classical states, when relative angle uncertainties are small. From an operational standpoint, to model the corrections we need coherent states that are $\delta$-classical - one condition for each angle. Scattering data can be used to compare with the expected corrections from the coherent states to determine the agreement with classical 3-dimensional angular geometry on the effective scale. The consistency of the angular data can be used to check the homogeneity assumption.  The same method can be used to check the effective dimensionality of mesoscopic spatial geometry. Specifying the only angular data for such states then reduces the requirements on the states and thus the combinatorics may provide a lever arm similar to the model reviewed above. 

\end{document}